# The hybrid magnet with a logarithmic in time field deviation.

E.P.Krasnoperov[1], A.A.Kartamyshev[1], D.I. Puzanov[1],
O.L.Polushchenko[2], N.A.Nizelskij[2]
[1]*Kurchanov Institute, 123182, Moscow, kep@isssph.kiae.ru*
[2] *Bauman MSTU, 105005, Moscow*

*The pulsed-field magnetization of hybrid magnet (soft ferromagnetic and melt-grown superconductor Y-Ba-Cu-O) was investigated at T = 78K. It is shown that choosing the pulse amplitude one can create a logarithmic in time field deviation near the specified value.*

High-temperature superconductors (HTS) can to trap magnetic field due to its high super current density - $J_c$ [1]. In superconductors such as melt- grown type of Y-Ba-Cu-O super current density at liquid nitrogen temperature (T = 78K, H=1T) reaches $J_c$ = 10 kA/cm$^2$, and in a twinned single crystals $J_c$ is 3 times higher [2]. On the basis of such materials one may create a permanent magnet with a large field, which are of interest to compact NMR [3], for magnetron sputtering device [4] for the sources of synchrotron radiation, as electron beam control elements [5] and other purposes. To obtain a homogeneous magnetic field it is convenient to use superconducting ring. According to Bean model of the critical state (See in [6]) current density in a superconductor is the critical one $J_c$ =10/4π·dB/dR. The field B is relatively uniform inside a ring's hole and is almost linearly decreases from the inner surface of the ring to the outside [6,7]. This refers to the condition of slow (isothermal) magnetization, when the local heating and delay of the field penetration can be neglected. The longitudinal magnetic field in the center of the ring having a square cross section and with an inner radius equal to the width (in generally accepted notation h=a=b/2 [8]) is H≈0.4π·I·ln2. Here I is a ring's current. Current density $J_c$ = 10 kA/cm$^2$ in the ring having cross section of 1 cm$^2$ can generate magnetic field H~1T. We propose to use superconducting ring for the magnetization of soft ferromagnetic as permendyur (Fe-Co with 2%V) that has a saturation induction about 2.3 T. Then such hybrid magnet can create a homogeneous field of up to 3 T. This is considerably higher than that of the known hard materials such as Nd-Fe-B. Ferromagnetic core in hybrid magnet is used to enhance the field and achieve a high homogeneity. Superconducting ring supports continuous core magnetization. With decreasing temperature down to 68K the super current density in a superconductor increases more than 2 times. One can expect the emergence of hybrid permanent magnets with a homogeneous field of 4-5 T working in nitrogen temperature.

From practical point of view a pulse field magnetization (PFM) is a well - situated to magnetically activate the superconductor. This method is wide used to magnetize a HTS tablets [9,10]. Pulsed magnetization of multiply connected superconductors is of particular scientific and practical interest. A penetration field time into a superconductor at a distance - l is $\tau = l^2/D$, where $D_m = c^2/4\pi\rho$ - diffusion coefficient of the field, ρ- resistively of vortex motion. If the pulse duration $\tau_i$ is shorter than τ, then the dynamics of vortex and field trap are significantly depend on effect of field penetration delay and local heating. Perhaps this explains the unusual distribution of the fields in the drilled tablet [11]. In our work characteristics of a hybrid magnet under a short pulse magnetization is examined.

The design of the magnet is shown schematically in Fig. 1. Two permendyur cylinders of a diameter 16 mm and a 20 mm of length are located in the coaxial gap 1.5 mm. Each cylinder is seized by ring of HTS melt- grown Y-Ba-Cu-O outside diameter of 38 mm and 11 mm of thickness. Rings cut from discs that were grown using an elongated seed [12]. Studied system is immersed in liquid nitrogen at T = 78K. Magnetization was realized by pulsed coil, which generates a field close to half the period of 10 ms sinusoids. It is important to emphasize that a pulsed field direction remained unchanged. The interval between magnetizing pulses was 1-3 min. This wait time between the pulses is chosen for the need to complete the cooling of superconducting rings after pulse heating. In the center of the magnet gap the Hall sensor is set to measure local field. Two devices: a digital oscilloscope and DC voltmeter were connected to Hall sensor in parallel. Oscilloscope measured instantaneous field in the gap, and simultaneously the magnetizing pulse current with the help of Rogovskii coil. DC voltmeter is used to determine the long-term evolution of field captured at intervals of 1-10 sec.

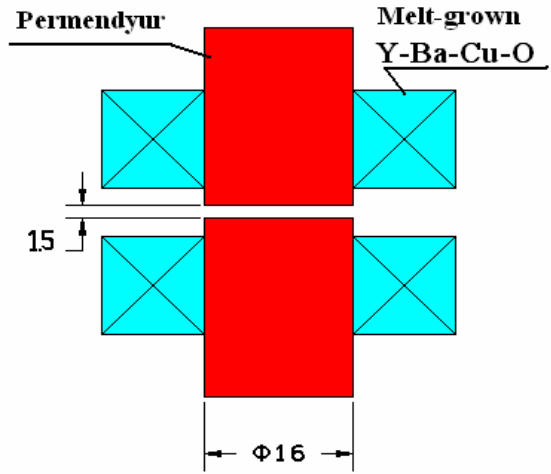

Fig.1. Design of the hybrid magnet.

Experiments have shown that the pulse magnetization of superconducting ring is essentially different from the disc. In case of disk at T=78K single magnetizing pulse with amplitude $\mu H_a$ = 2,7 T can create a field ≈1T [10,13]. In the ring the fields with amplitudes > 1.5 T destroy ring supercurrent but large the ring's body magnetization remains (as in ferromagnetic). Thus flux in ring's hole is changed on the opposite direction and field in this case reaches ≈0,6 T. Similar effect was observed in tubes of Bi-2212 [14].

The typical magnetization curves of a hybrid magnet in field near B≈1.3 T for various pulse amplitudes are shown in the inset in Fig.2. After magnetizing pulse the jump of trap flux is observed. And then field in gap is increased. It is seen that the jumps can be either upward field (the first 2 jumps), and to its decline (3rd race). As ΔB conditional difference field has been chosen to jump and its value in 2 seconds after the pulse. Fig. 2 shows the dependence of ΔB on the amplitude of the momentum in the field B≈1,3 T. If pulse amplitudes are higher µH$_a$ > 1.1T the field in the gap increases (ΔB>0), but at lower amplitudes, the field reduces. If amplitudes are near 1.1 T the field practically does not change.

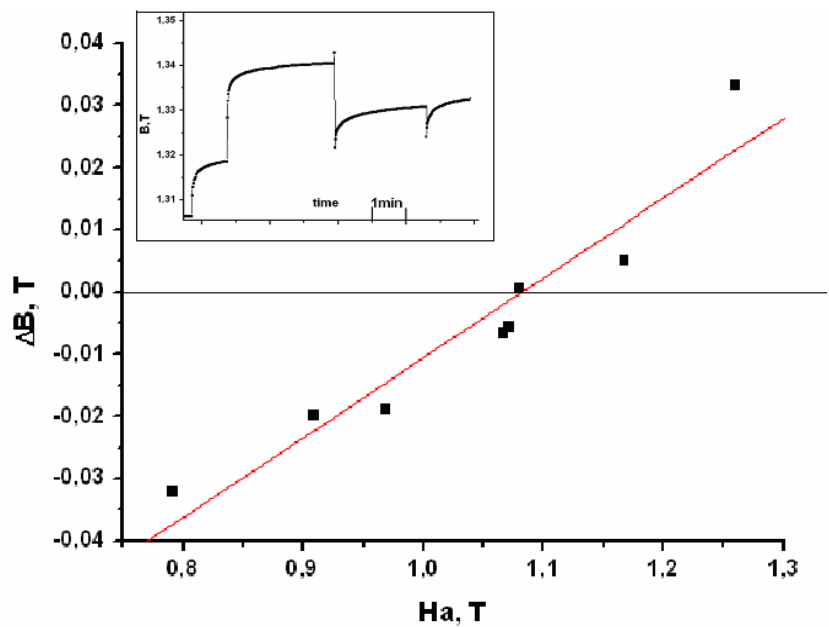

Fig.2. The trap flux jumps ΔB vs. magnetic field pulse amplitude.

After magnetizing pulse, the field increases with time, regardless of whether a leap up or down. This is magnetic flux relaxation that is well known phenomenon in the superconductor [15]. As in the case of the tablet a growth of the field (negative relaxation in the conventional definition [13]) is provided that the maximum-trapped field is located inside a ring's body of R$_{out}$ <R$_{max}$ <R$_{in}$. In this case, supercurrents in the inside part of the ring (J$_{in}$) flow due to the increasing branch of the pulse. In the outside part of ring supercurrents (J$_{out}$) flow in opposite direction, due to fall of the pulse. When pulse exposures, the outer part of the ring is heated more strongly than the inner. Since supercurrent reduces with increasing temperature (dJ$_c$ /dT <0), the outside supercurrent is smaller than the inner one (J$_{out}$<J$_{in}$). After cooling down to the equilibrium temperature (78K) J$_{out}$ relaxes slower in compare to J$_{in}$ [13]. As a result of more rapid decay of internal supercurrent the field in the center of the ring grows and consequently the core magnetization increases. As shown in Fig.2 the field in the gap can be reduced by magnetization pulse. Selecting pulse amplitude one can reduce B value on amount of relaxation growth and thereby create periodic deviation fields: relaxation growth and a sharp reduction by the magnetizing pulse. An example of such a regime for the field B≈0.74 T is shown in Fig. 3. Time delayed in the logarithmic scale, and the countdown begins once again, after each pulse.

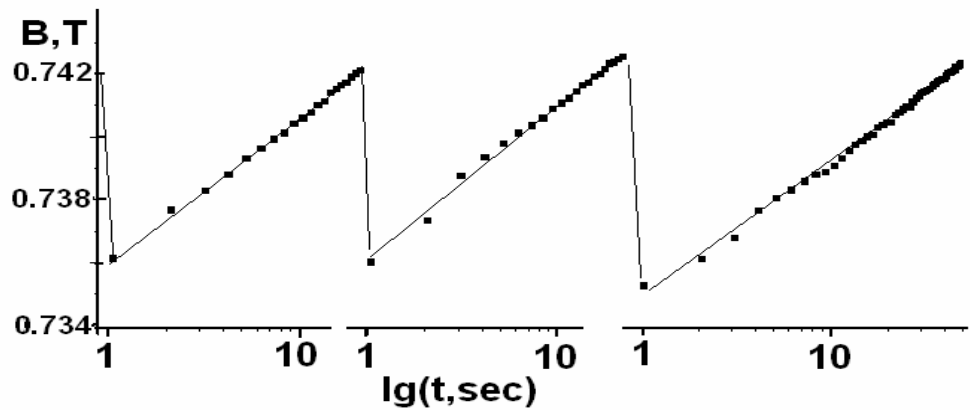

Fig.3. Deviation of trapped flux after pulse magnetization.

The amplitude of pulses is µH$_a$≈1.0 T, and they follow through 30-40 seconds. Since the relaxation is linear in logarithmic scale of time, the field increases linearly in this scale. Magnetizing pulse applied after 30 seconds restores the original image structure. Field has filed, and then again increases. Thus is realized a field's deviation with linear logarithmic time scale.

With increasing of the magnetization, ring's area of internal supercurrent J$_{in}$ is reduced. Near the maximum magnetization B≈2.3T field deviation is also released. But in this case pulse with amplitude H$_a$ ≈1.35T produces field jump in up and after that field goes down linearly in logarithmic time scale. The results of studies of supercurrent relaxation in the superconducting ring will be published later.

As a result by short pulses one can control the magnetization melt-grown HTSC ring and create on their basis the hybrid magnet with logarithmic in time deviations field.